\documentclass[aip,apl,reprint,superscriptaddress]{revtex4-1}

\usepackage{graphicx}
\usepackage{SIunits}
\usepackage{color}

\begin{document}

\title{Switching Nonlinearity in a Superconductor-Enhanced Metamaterial}
\author{Cihan Kurter}
\affiliation{Center for Nanophysics and Advanced Materials, Department of Physics, University of Maryland, College Park, Maryland 20742-4111, USA}

\author{Philippe Tassin}
\affiliation{Ames Laboratory---U.S. DOE and Department of Physics and Astronomy, Iowa State University, Ames, IA 50011, USA}
\affiliation{Applied Physics Research Group, Vrije Universiteit Brussel, Pleinlaan 2, B-1050 Brussel, Belgium}

\author{Alexander P. Zhuravel}
\affiliation{B. Verkin Institute for Low Temperature Physics and Engineering, National Academy of Sciences of Ukraine, 61103 Kharkov, Ukraine}

\author{Lei Zhang}
\affiliation{Ames Laboratory---U.S. DOE and Department of Physics and Astronomy, Iowa State University, Ames, IA 50011, USA}

\author{Thomas Koschny}
\affiliation{Ames Laboratory---U.S. DOE and Department of Physics and Astronomy, Iowa State University, Ames, IA 50011, USA}

\author{Alexey V. Ustinov}
\affiliation{Physikalisches Institut and DFG-Center for Functional Nanostructures (CFN), Karlsruhe Institute of Technology, D-76128 Karlsruhe, Germany}

\author{Costas M. Soukoulis}
\affiliation{Ames Laboratory---U.S. DOE and Department of Physics and Astronomy, Iowa State University, Ames, IA 50011, USA}
\affiliation{Institute of Electronic Structure and Lasers (IESL), FORTH, 71110 Heraklion, Crete, Greece}

\author{Steven M. Anlage}
\affiliation{Center for Nanophysics and Advanced Materials, Department of Physics, University of Maryland, College Park, Maryland 20742-4111, USA}
\affiliation{Physikalisches Institut and DFG-Center for Functional Nanostructures (CFN), Karlsruhe Institute of Technology, D-76128 Karlsruhe, Germany}

\date{\today}

\begin{abstract}
We demonstrate a nonlinear metamaterial that can be switched between low and high transmission by controlling the power level of the incident beam. The origin of this nonlinear response is the superconducting Nb thin film employed in the metamaterial structure. We show that with moderate RF power of about \unit{22}{\textrm{dBm}} it is possible to quench the superconducting state as a result of extremely strong current densities at the corners of the metamaterial's split-ring resonators. We measure a transmission contrast of \unit{10}{\deci\bel} and a change in group delay of \unit{70}{\nano\second} between the low and high power states.
\end{abstract}

\maketitle

The field of metamaterials---or artificial materials in which small electric circuits replace atoms as the basic unit of interaction with electromagnetic radiation---has made remarkable progress in recent years.\cite{Smith-2004,Soukoulis-2010} Electromagnetic metamaterials with exotic properties such as magnetic response at terahertz and optical frequencies, negative index of refraction, and giant chirality have been developed,\cite{Shalaev-2005,Soukoulis-2010,Plum-2009} with potential applications for super-resolution lenses,\cite{Pendry-2000} subwavelength and terahertz photonics,\cite{Engheta-2007} and transformation optics.\cite{LeonhardtPendry} Most metamaterials designed to date show linear electromagnetic response, sometimes tunable by external control through temperature, mechanical adjustment, or photocarriers in semiconductors.\cite{tunable} Despite a wide range of potential applications,\cite{nonlinearmm} nonlinear metamaterials have turned out to be much harder to design. At microwave frequencies, it is possible to create nonlinear electromagnetic response using nonlinear lumped elements, e.g., varactors or diodes,\cite{lumped} but this approach becomes unfeasible at higher frequencies.

In this Letter, we demonstrate a nonlinear metamaterial without using lumped elements. This metamaterial (see inset in Fig.~1) contains a \unit{200}{\nano\meter}-thick split-ring resonator (SRR) pair made from Nb symmetrically located around a \unit{2}{\micro\meter}-thick cut wire made from Cu on a quartz substrate; the details of the fabrication technique can be found in Ref.~12. The wire couples directly to the incident electromagnetic waves (radiative element), while the SRRs have vanishing dipole interaction with the excitation field (dark element). We have shown previously that this design allows for an electromagnetic response analogous to media exhibiting electromagnetically induced transparency (EIT)---a narrow transparency window with large group delay.~\cite{EIT}  A high transmission contrast is produced by (i) using the antisymmetric hybridization of the magnetic dipole resonances of the SRRs as the dark mode to avoid magnetic dipole radiation, and (ii) using a superconductor in the SRRs to maximize the loss contrast between the dark and radiative resonators at temperatures below the critical temperature, $T_\mathrm{c}$, of Nb.\cite{UM-ISU}

\begin{figure}[b]
\centering
\includegraphics[clip,width=2.5in]{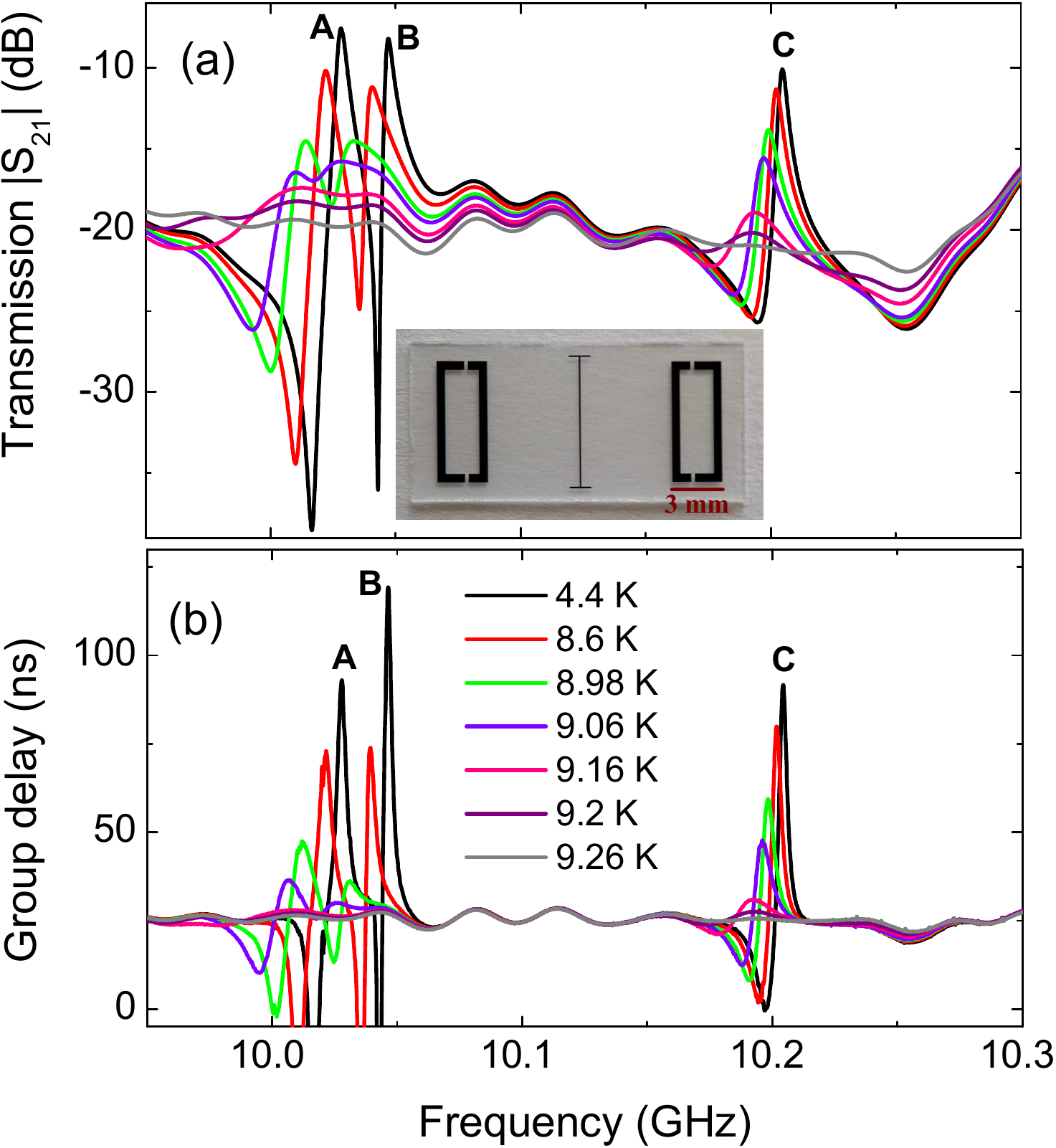}
\caption{(Color online) (a)~Transmission $|S_{21}|$ and (b)~group delay spectra for the metamolecule mounted in an X-band waveguide for a set of temperatures. The inset in (a) shows a photograph of the sample with the cut wire and the SRR pair. The labels A, B, and C are used to identify the three EIT resonances.}
\label{Fig1}
\end{figure}

We first show that introducing superconducting Nb thin films into the dark elements allows for an effective and sensitive tuning of the RF response of the metamaterial through changes in the superfluid (Cooper pair) density, $n_\textrm{s}$, which decreases with increasing temperature and drops to zero at the critical temperature $T_\textrm{c}$.\cite{Tinkham,KurterAPL} The quartz substrate carrying the superconducting SRRs and the copper cut wire is placed in the center of the crosssection of a Nb WR-90 waveguide illuminated with the fundamental $\mathrm{TE}_{10}$ mode. After cooling down below the critical temperature of Nb, we measure the reflection and transmission amplitudes using a vector network analyzer. In Fig.~\ref{Fig1}(a)-(b), we plot the transmission, $|S_{21}|$, and group delay as a function of frequency for a set of temperatures at fixed RF input power ($\unit{-10}{\mathrm{dBm}}$, corresponding to a power density of \unit{50}{\micro\watt\per\centi\meter\squared}). At \unit{4.4}{\kelvin}, we observe three intense EIT-like resonant peaks in transmission. Each transparency window is accompanied by a large
group delay of the order of \unit{100}{\nano\second}. The enhancement in both transmission and group delay for each dark mode weakens with increasing temperature because of the increase in ohmic loss of the dark element.\cite{Ricci} Finally, at $T_c \approx \unit{9.26}{\kelvin}$, Nb becomes a normal metal resulting in closure of the transparency windows and no enhancement in group delay. Since the EIT resonances of the metamaterial under study rely critically on the loss difference between dark and radiative elements, the superfluid density $n_\textrm{s}$ in the Nb film acts as a powerful knob to modulate the response.

\begin{figure}
\centering
\includegraphics{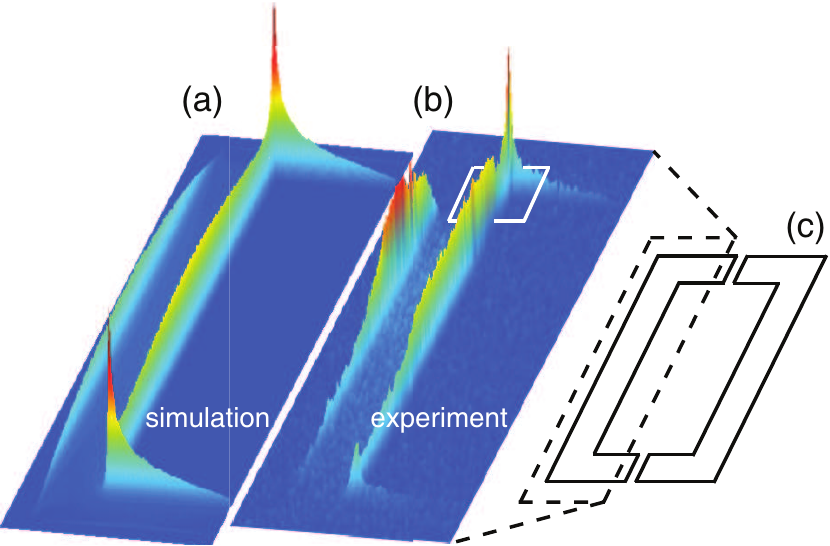}
\caption{(Color online) Current densities in one arm of the SRR. (a)~Results obtained using a frequency-domain electromagnetics solver. (b)~Experimental LSM photo-response image. The LSM image was acquired at a frequency of \unit{9.747}{\giga\hertz}, input power of \unit{+18}{\textrm{dBm}}, and a temperature of \unit{7}{\kelvin}. (c)~Schematic of the leftmost SRR of the sample. The dashed box indicates the area of the SRR in which the current densities are shown in (a) and (b).}
\label{Fig2}
\end{figure}

The above experiment still requires an exter\-nal variable---the temperature---to modify the metamaterial's response.\cite{Chen-2010} As soon as we introduce a superconducting material, one could expect that the metamaterial's response becomes nonlinear, because of two effects. (1) The superfluid density depends on the magnetic field, as well as the local current density. (2) The superfluid density depends on the temperature, which can be a function of the local current density. Therefore, if the incident microwave power is increased and, hence, the current densities become larger, these effects should appear. In earlier works, we were unable to achieve sufficiently high power densities to reach the nonlinear regime. Recently, we realized that by deliberately making SRRs with sharp square interior corners (which is normally avoided because of higher loss~\cite{Guney-2009}), we could create highly localized spots of enhanced current density. If the enhanced currents at the corners can reach the critical current density of the superconductor, the nonlinear regime might become accessible with the moderate microwave power levels of our experimental setup. This idea is confirmed by the results of our finite-element simulations (using CST Microwave Studio) shown in Fig.~\ref{Fig2}(a). We observe large current densities flowing at the interior and exterior edges of the Nb thin film SRR. The microwave current density is enhanced at the edges to screen the interior of the SRRs from the electromagnetic fields.\cite{ZhuravelAPL} At the interior corners, the currents are squeezed through a thinner path, leading to very large peaks in the current density (in our simulations, the peak values are limited by the discretization). 

This phenomenon is confirmed by our current density measurements using laser scanning microscopy (LSM). The LSM technique images the photoresponse, $PR \sim (\partial |S_{11}|/\partial T) \delta T$, where $|S_{11}|$ is the reflection amplitude and $\delta T$ is the local temperature change due to laser heating. This response is essentially resistive in nature and proportional to the local value of microwave current density squared, $J_{RF}^2(x,y)$.\cite{RicciIEEE,ZhuravelLTP} The LSM images of the SRR plotted in Fig.~\ref{Fig2}(b) are in excellent agreement with the simulations.

\begin{figure}
\centering
\includegraphics[bb=90 187 518 644,width=2.5 in]{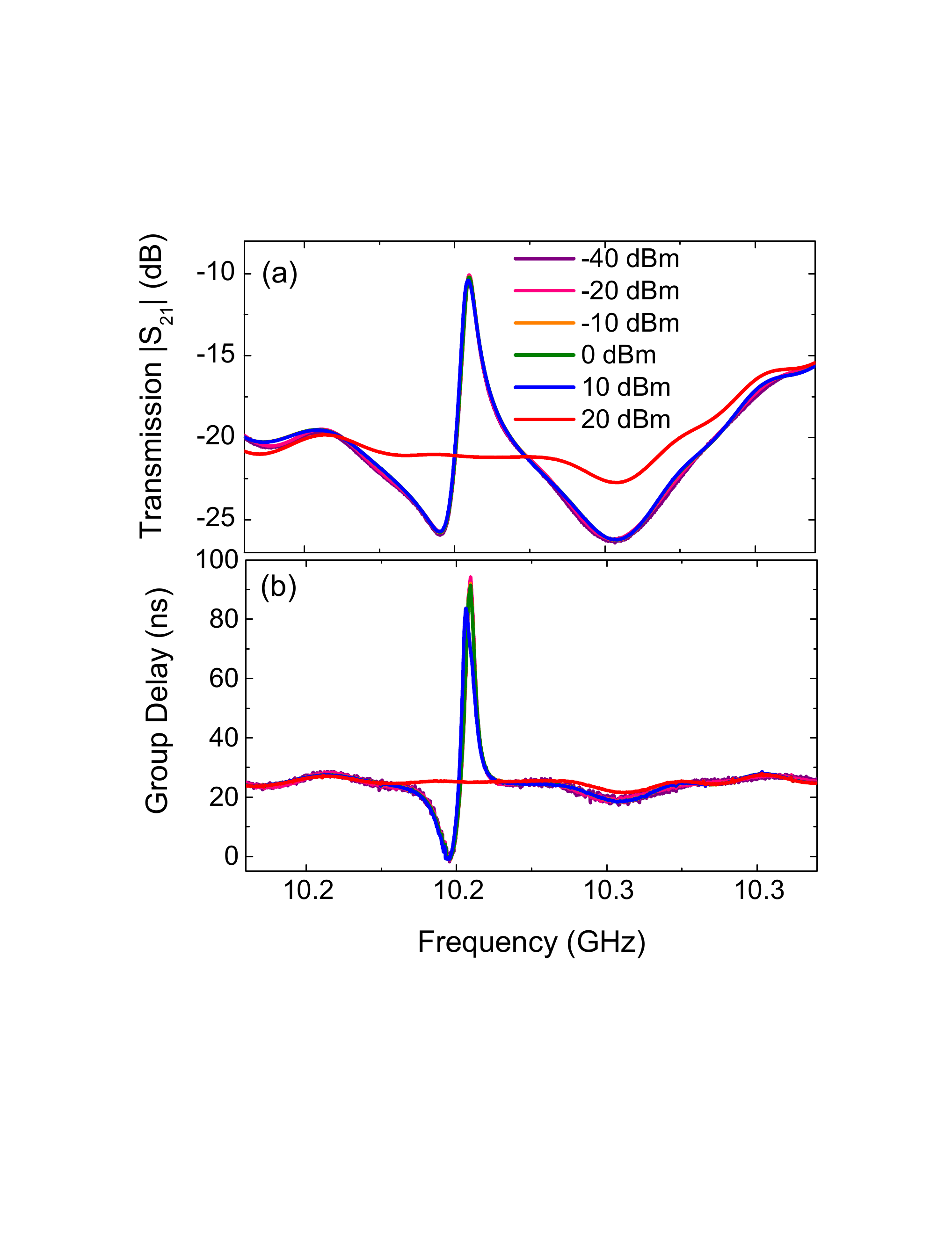}
\caption{(Color online) Input power dependence of (a)~the transmission $|S_{21}|$ and (b)~ the group delay in the transmission window labeled \textit{C} in Fig.~\ref{Fig1}.}
\label{Fig3}
\end{figure}

Subsequently, we have examined the microwave power dependence of transmission and group delay of the EIT feature labeled \textit{C} in Fig.~\ref{Fig1} at \unit{4.4}{\kelvin}. The results are plotted in Fig.~\ref{Fig3}(a)-(b). We observe high transmission and large group delay for input powers below \unit{20}{\textrm{dBm}}, and a sharp switching event that suddenly extinguishes the EIT transmission and group delay enhancement for higher power levels. There is no gradual change as in the temperature control experiment described above. When the input power is increased, the current densities in the SRRs become larger, but not large enough to heat the superconductor significantly. However, when the large peaks in the current density at the interior corners reach the critical value, the superconducting state is quenched. Thus we have created a nonlinear switchable metamaterial in which the slowdown of electromagnetic radiation can be turned on or off simply by modulation of the incident power. (Note that we have observed the same switching behavior for the peaks labeled \textit{A} and \textit{B}, but at slightly larger input powers. This is because the three EIT resonances have different current distributions in the SRRs (different dark modes) and, therefore, different peak values in the corners.)

\begin{figure}
\centering
\includegraphics[bb=94 349 508 659,width=3 in]{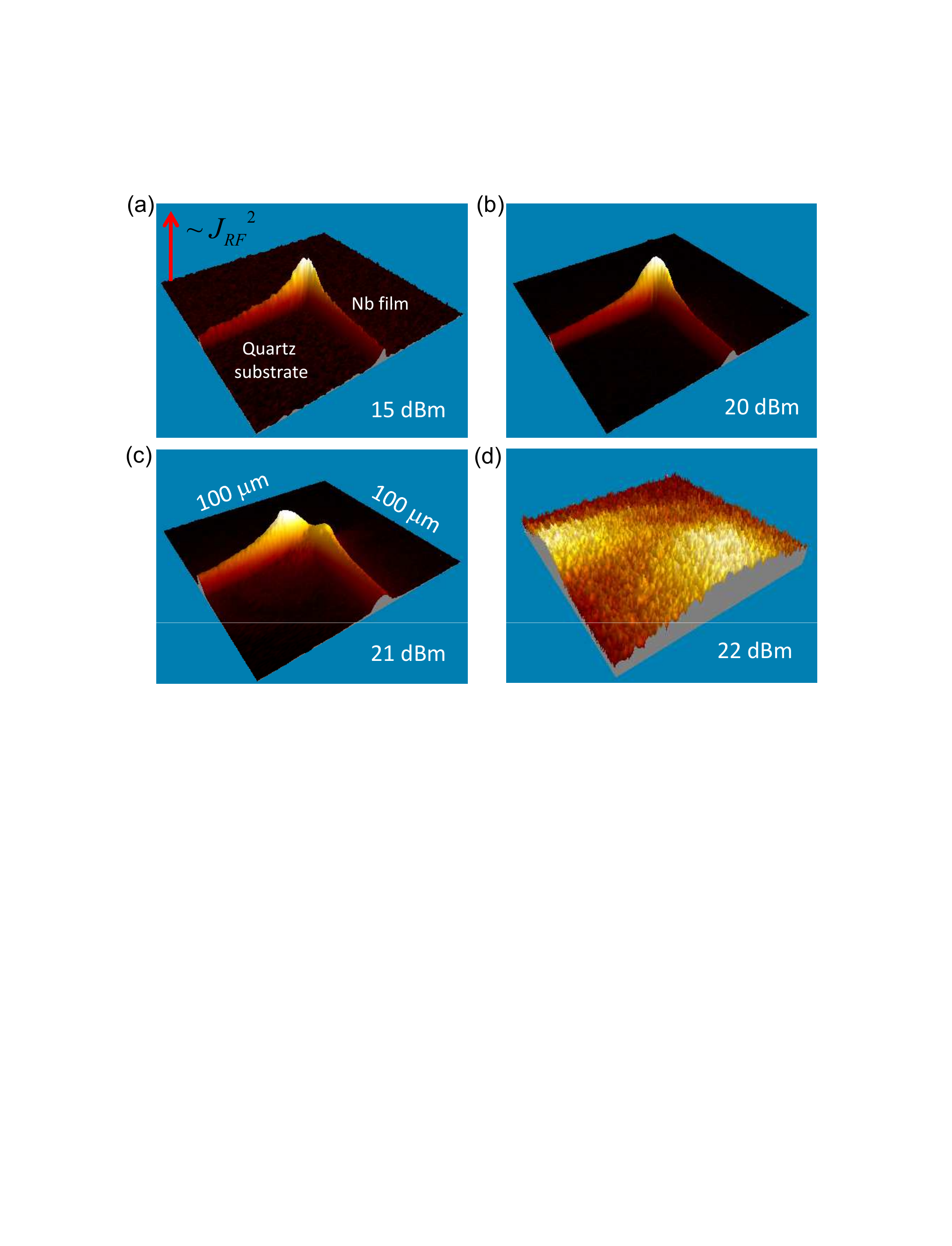}
\caption{(Color online) Images of the LSM photoresponse in the vicinity of the interior square corner of the Nb SRR [white square in
Fig.~\ref{Fig2}(b)] as a function of incident microwave power from \unit{+15}{\textrm{dBm}} to \unit{+22}{\textrm{dBm}}. The images cover an area of $\unit{100}{\micro\meter}\times\unit{100}{\micro\meter}$. (Note that the EIT sample is in a different configuration while measured with the LSM technique compared to the waveguide measurements shown in Figs.~\ref{Fig1} and~\ref{Fig3}. The incident field configurations will therefore be somewhat different, resulting in slightly different critical input powers at which the transparency window is destroyed in the two experiments.)}
\label{Fig4}
\end{figure}

Since we believe the current densities at the interior corners of the Nb SRRs to be of critical importance to the switching process, we have repeated the LSM photoresponse experiments in a $\unit{100}{\micro\meter}\times\unit{100}{\micro\meter}$ area at a lithographically defined interior corner of the Nb SRR. In Fig.~\ref{Fig4}, we see that the photoresponse peak emerging at the interior corner of the Nb SRR increases in magnitude and spreads out, suggesting the development of a critical state.\cite{RicciIEEE,ZhuravelIEEE} In that state, there is magnetic flux generated by the microwave currents that enters the
Nb film and causes dissipation. As the critical state extends further into the material, the dissipation leads to thermal runaway and eventually to quenching of the superconductivity at \unit{+22}{\textrm{dBm}}. This in turn leads to a decrease in the resonator's quality factor and switching off of the EIT features. Nevertheless, those features are fully recovered upon reduction of the microwave power below the switching value, i.e., there is no measurable hysteresis. By modeling the SRR as a microwave stripline excited by the fundamental waveguide mode, we estimate the peak current density at the corner to be roughly  $\unit{4 \times 10^{11}}{\ampere\per\meter\squared}$. This is slightly below the critical depairing current of Nb calculated from the Ginzburg-Landau theory ($J_\mathrm{c} = \unit{1.75 \times 10^{12}}{\ampere\per\meter\squared}$), but this can be attributed to an underestimation of the peak current density at the corner or to the critical depairing current being smaller due to vortex entry/flow and/or heating of the sample.

In summary, we have demonstrated a precise and sensitive tuning of a superconductor/metal hybrid metamaterial exhibiting classical EIT resonances through the modulation of superfluid density in the superconductor. We have observed a sharp switching event of the transparency windows at high microwave powers, resulting in a nonlinear switchable metamaterial without the use of varactors or other lumped elements.

The work at Maryland was supported by ONR Award No.\ N00014-08-1-1058 and 20101144225000, the U.S.\ DOE (High Energy Physics) under Contract No.\ DESC0004950, the ONR/UMD AppEl Center, task D10 (Award No.\ N00014-09-1-1190), and CNAM.  The work at Ames Laboratory was partially supported by the U.S. DOE, Basic Energy Science, Division of Materials Sciences and Engineering, under Contract No.\ DE-AC02-07CH11358 (computational studies), by ONR Award No.\ N00014-10-1-0925 (characterization), and by the European Community FET project PHOME, Contract No.\ 213310 (theoretical studies). The work in Karls\-ruhe and Kharkov is supported by a NASU program on ``nanostructures, materials and technologies.'' S.M.A. acknowledges sabbatical support from the DFG-Center for Functional Nanostructures at KIT, and P.T. acknowledges a fellowship from the Belgian American Educational Foundation.

\end{document}